\documentclass[prd]{revtex4}
\usepackage{graphics}
\usepackage{epsfig}

\newcommand\ba{\begin{eqnarray}}
\newcommand\ea{\end{eqnarray}}
\newcommand\nn{\nonumber}

\newcommand{\br}[1]{\left( #1 \right)}
\newcommand{\brs}[1]{\left[ #1 \right]}
\newcommand{\brf}[1]{\left\{ #1 \right\}}
\newcommand{\brm}[1]{\left| #1 \right|}

\newcommand{\GeV}{~\mbox{GeV}}
\newcommand{\MeV}{~\mbox{MeV}}
\newcommand{\KeV}{~\mbox{keV}}

\begin{document}

\title{Radiative decays of scalar mesons $\sigma(600)$, $f_0(980)$ and $a_0(980)$
in the local Nambu-Jona-Lasinio model}

\author{\bf\firstname{M.~K.}~\surname{Volkov}}
\email{volkov@theor.jinr.ru}
\affiliation{%
Joint Institute for Nuclear Research, Dubna, Russia
}%
\author{\bf\firstname{E.~A.}~\surname{Kuraev}}
\email{kuraev@theor.jinr.ru}
\affiliation{%
Joint Institute for Nuclear Research, Dubna, Russia
}%

\author{\bf\firstname{Yu.~M.}~\surname{Bystritskiy}}
\email{bystr@theor.jinr.ru}
\affiliation{%
Joint Institute for Nuclear Research, Dubna, Russia
}%

\begin{abstract}
The two-photon decay widths of scalar mesons
$\sigma(600)$, $f_0(980)$ and $a_0(980)$ as well
$a_0\to \rho(\omega)\gamma$ and $f_0\to \rho(\omega)\gamma$
are calculated in the framework of
the local Nambu-Jona-Lasinio model. The contributions
of the quark loops (Hartree-Fock approximation) and the meson
loops (next $1/N_c$-approximation where $N_c$ is the number of colors)
are taken into account.
These contributions, as we show, are the values of the same order of magnitude.
For the $f_0$ decay the $K$-loop contribution turns out to play the
dominant role.
The results for two-gamma decays are in satisfactory agreement with modern experimental
data.
The predictions for
$a_0\to \rho(\omega)\gamma$ and $f_0\to \rho(\omega)\gamma$
widths are given.
\end{abstract}

\maketitle


\section{Introduction}

In recent paper \cite{Bystritskiy:2007wq}, the radiative decay
$\phi\to f_0\gamma$ width
within the local Nambu-Jona-Lasinio (NJL) model
\cite{Volkov:1986zb,Volkov:1982zx,Ebert:1982pk,Volkov:1984kq,Ebert:1985kz,Ebert:1994mf,Volkov:2006vq,Vogl:1991qt,Klevansky:1992qe}
has been calculated.
In this work, same as in \cite{Bystritskiy:2007wq},
we took into account not only
the quark loop contributions (Hartree-Fock approximation)
but also the meson loop contributions
(next $1/N_c$-approximation where $N_c$ is the number of colors).
We should note that meson loops
give the contribution of the same order as the Hartree-Fock approximation
due to fractional charge of quarks and integer charge of mesons.
The gauge invariance of the amplitudes
leads to absence of ultraviolet divergences in
the relevant loop integrals.
Thus, the explicit dependence of these amplitudes of the external momenta
was obtained \cite{Ebert:1996pc}.

In this paper, we will consider the two-photon decays of the scalar mesons
$\sigma(600)$, $f_0(980)$ and $a_0(980)$ as well as radiative decays
$a_0\to \rho(\omega)\gamma$ and $f_0\to \rho(\omega)\gamma$.

In the case of the quark loop we consider only a real part of the
relevant loop integral.
This prescription permits us to take into account the condition of
the "naive" quark confinement. Some theoretical arguments
supporting this procedure can be found in \cite{Pervushin:1985yi}.
As for the meson loops, both the real and the imaginary parts were
taken into account.

The structure of our paper is the following.
In Section~\ref{Lagrangian}, the NJL quark-meson Lagrangians,
corresponding parameters and the coupling constants of our model
are defined.
In Section~\ref{LoopIntegrals}, the methods of
quark and meson loop calculation are described;
the contributions of quark and meson loops
to the amplitudes and the widths of
two-photon decays of the scalar meson are presented.

In Section~\ref{ChapterNew}, we consider the
$a_0\to \rho(\omega)\gamma$ and $f_0\to \rho(\omega)\gamma$
decay widths.

In Section~\ref{Conclusion}, we discuss the results obtained.

\section{Lagrangian of the NJL model}
\label{Lagrangian}

The initial four-quark Lagrangian of the local Nambu-Jona-Lasinio model
has the form \cite{Volkov:1986zb,Ebert:1985kz,Vogl:1991qt,Klevansky:1992qe}:
\ba
    {\cal L} &=&
    \bar q \brs{ i \hat\partial + e Q \hat A - M_0 }q
    +
    \nn\\
    &+&
    \frac{G}{2}\br{ \br{\bar q q}^2 + \br{\bar q \br{i\lambda^a \gamma_5}q}^2}
    -
    \frac{G_V}{2}\br{ \br{\bar q \gamma_\mu \lambda^a q}^2 + \br{\bar q \br{\gamma_\mu \gamma_5 \lambda^a }q}^2},
\ea
where $\bar q= \br{\bar u,\bar d,\bar s}$, and $u$, $d$, $s$ are the quark fields,
$M_0=\mbox{diag}\br{m^0_u,m^0_d,m^0_s}$ is the matrix of quarks current masses,
$Q=\mbox{diag}\br{2/3,-1/3,-1/3}$ is the quark electric charge matrix,
$e$ is the elementary electric charge ($e^2/4\pi=\alpha=1/137$),
$\lambda_i$, $i=1\dots 8$ are the well-known Gell-Mann matrices and
$\lambda_0 = \sqrt{2/3}~\mbox{diag}\br{1,1,1}$,
$G$ and $G_V$ are effective coupling constants of four-fermion interactions.

The procedure of bosonization and the renormalization of meson fields
leads to the effective
quark-meson Lagrangian. The part of this Lagrangian which we will use
is the following \cite{Volkov:1986zb,Ebert:1994mf,Volkov:2006vq}:
\ba
    {\cal L}_{eff} &=&
    \bar q \brs{ i \hat\partial + e Q \hat A - M }q
    +
    \bar q \left[\frac{}{}
    g_{\sigma_u} \lambda_u \sigma_u + g_{\sigma_s} \lambda_s \sigma_s + g_{\sigma_u} \lambda_3 a_0 +
    i \gamma_5 g_\pi \br{\lambda_{\pi^+} \pi^+ + \lambda_{\pi^-} \pi^-}
    +\right. \nn\\
    &&\qquad+\left.
        i \gamma_5 g_K \br{\lambda_{K^+} K^+ + \lambda_{K^-} K^-}
        +
        \frac{g_\rho}{2} \br{ \lambda_3 \hat \rho_0 + \lambda_u \hat \omega }
    \right] q,
    \label{QuarkMesonLagrangian}
\ea%
where $M=\mbox{diag}\br{m_u,m_d,m_s}$ is the constituent quark mass matrix
and $m_u=m_d=263\MeV$, $m_s=406\MeV$,
$\lambda_u=\br{\sqrt{2} \lambda_0+\lambda_8}/\sqrt{3}$,
$\lambda_s=\br{-\lambda_0+\sqrt{2}\lambda_8}/\sqrt{3}$,
$\lambda_{\pi^\pm}=(\lambda_1\pm i\lambda_2)/\sqrt{2}$,
$\lambda_{K^\pm}=(\lambda_4\pm i\lambda_5)/\sqrt{2}$.
Taking into account the six-quark interaction of t'Hooft leads
to mixing of $\sigma_u$ and $\sigma_s$ states
\cite{Volkov:2006vq,Vogl:1991qt,Klevansky:1992qe,Ebert:1994mf}.
Scalar isoscalar mesons $f_0$, $\sigma$ are the mixed states
\ba
f_0 &=& \sigma_u\sin\alpha+\sigma_s\cos\alpha, \nn \\
\sigma &=& \sigma_u\cos\alpha-\sigma_s\sin\alpha,
\label{Mixing}
\ea
with the mixing angle $\alpha=11.3^o$ \cite{Volkov:2006vq,Vogl:1991qt}.

The coupling constants from the Lagrangian (\ref{QuarkMesonLagrangian}) are
defined in the following way \cite{Volkov:1986zb}:
\ba
g_{\sigma_u} &=& \left( 4 I^\Lambda\br{m_u, m_u}\right)^{-1/2} = 2.43, \nn\\
g_{\sigma_s} &=& \left( 4 I^\Lambda\br{m_s, m_s}\right)^{-1/2} = 2.99, \nn \\
g_{\pi} &=& \frac{m_u}{F_\pi} = 2.84, \nn\\
g_K &=& \frac{m_u+m_s}{2 F_K} = 3.01, \nn \\
g_\rho &=& \sqrt{6} g_{\sigma_u} = 5.95, \nn
\ea
where we use the Goldberger-Treiman relation for $g_{\pi}$ and $g_K$ constants,
$F_\pi=92.5\MeV$ and $F_K = 1.2~F_\pi$, and $I^\Lambda\br{m, m}$ is the
logarithmically divergent integral which has the form:
\ba
    I(m,m) &=& \frac{N_c}{\br{2\pi}^4}
    \int d^4 k
    \frac{\theta\br{\Lambda^2-k^2}}
    {\br{k^2+m^2}^2} = \nn\\
    &=&
    \frac{N_c}{\br{4\pi}^2}
    \br{
        \ln\br{\frac{\Lambda^2}{m^2}+1}
        -
        \frac{\Lambda^2}{\Lambda^2 + m^2}
    }, \qquad N_c = 3. \nn
\ea
This integral is written in the Euclidean space.
The cut-off parameter $\Lambda = 1.27\GeV$
is taken from \cite{Volkov:1986zb,Bystritskiy:2007wq}.

\section{Two-gamma decays of $\sigma(600)$, $f_0(980)$ and $a_0(980)$}
\label{LoopIntegrals}

The amplitudes of the $2\gamma$ decay can be expressed in terms of
the quark and meson loop integrals.

The quark loop contribution to the amplitude is given by
two triangle type Feynman diagrams (see Fig.~\ref{Fig1},~a):
\begin{figure}
\includegraphics[width=0.8\textwidth]{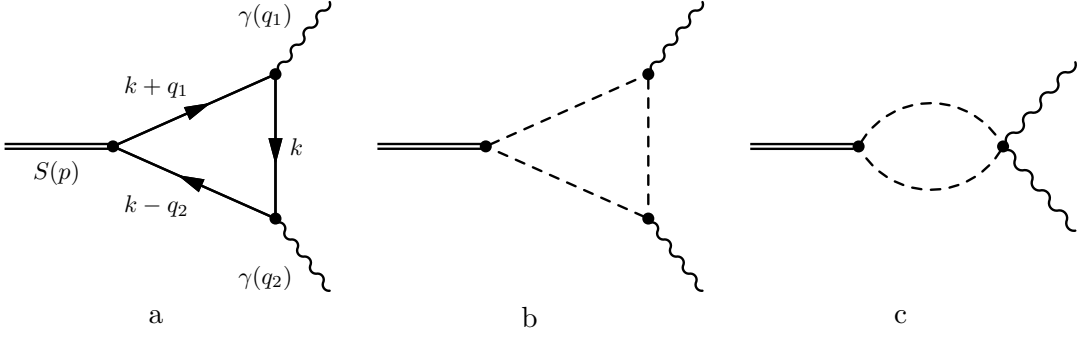}
\caption{The Feynman diagrams of quark and meson contributions
to two-photon decays of scalar mesons: $S \to 2\gamma$.
\label{Fig1}}
\end{figure}
\ba
    T_{\mu\nu}^q &=&
    -\frac{\alpha}{4\pi}
    \int \frac{d^4k}{i \pi^2}
    \frac{
        Sp\brs{
            \gamma_\nu\br{\hat k + m_q}\gamma_\mu\br{\hat k + \hat q_1 + m_q}\br{\hat k - \hat q_2 + m_q}
        }
    }
    {
        \br{k^2-m_q^2}
        \br{\br{k+q_1}^2-m_q^2}
        \br{\br{k-q_2}^2-m_q^2}
    }
    +\nn\\
    &+&\br{\br{q_1,\mu} \leftrightarrow \br{q_2,\nu}}.
    \ea
Applying the Feynman procedure of joining of the denominators
\ba
&&\frac{1}
    {
        \br{k^2-m_q^2}
        \br{\br{k+q_1}^2-m_q^2}
        \br{\br{k-q_2}^2-m_q^2}
    }
    = \nn\\
&&\qquad=
    \int\limits_0^1 dx
    \int\limits_0^1 2 y dy
    \frac{1}{\br{\br{k-q_x y}^2 - \br{m_q^2 + q_x^2 y^2}}^3},
\ea
where $q_x = x q_2 - \bar xq_1$, $\bar x = 1-x$, we obtain for
$T_{\mu\nu}^q$:
\ba
    T_{\mu\nu}^q &=&
        \frac{\alpha}{\pi}
        \br{g_{\mu\nu}\br{q_1 q_2} - q_{1\nu} q_{2\mu}} T^q, \\
    T^{q} &=& 2m_q
    \int\limits_0^1 dx
    \int\limits_0^1 y dy
    \frac{1-4y^2 x \bar x}{m_q^2 - M_S^2 y^2 x \bar x}.
\ea

For meson loops an additional Feynman diagram with two photon-two meson vertex
contributes as well
(see Fig.~\ref{Fig2},~b, c).
To restore the general gauge invariant form of the amplitude,
we can nevertheless consider only two triangle type Feynman diagrams:
\ba
    \Delta T_{\mu\nu}^M &=&
    \frac{\alpha}{4\pi}
    \int \frac{d^4k}{i \pi^2}
    \frac{
        \br{2k+q_1}_\mu \br{2k-q_2}_\nu
    }
    {
        \br{k^2-M^2}
        \br{\br{k+q_1}^2-M^2}
        \br{\br{k-q_2}^2-M^2}
    }
    +\nn\\
    &+&\br{\br{q_1,\mu} \leftrightarrow \br{q_2,\nu}}.
\ea
Extracting the term $\sim q_{1\nu}q_{2\mu}$ and adding the relevant term $\sim g_{\mu\nu}$
we obtain:

\ba
    T_{\mu\nu}^M &=&
        \frac{\alpha}{\pi}
        \br{g_{\mu\nu}\br{q_1 q_2} - q_{1\nu} q_{2\mu}} T^M,
\ea
with
\ba
 T^{M} &=& 2
    \int\limits_0^1 dx
    \int\limits_0^1 y dy
    \frac{y^2 x \bar x}{M^2 - M_S^2 y^2 x \bar x}.
\ea
Standard evaluation of these integrals leads to
\ba
    T^{q} &=& \frac{1}{m_q} F(z^q_S), \\
    T^{M} &=& \frac{z^M_S}{4M^2} \Phi(z^M_S),
\ea
where $z^q_S = 4 m_q^2/M_S^2$, $z^M_S = 4 M^2/M_S^2$,
\ba
    F(z) &=& \mbox{Re}\brs{1+\br{1-z}\Phi(z)}, \nn\\
    \Phi(z) &=& z \phi(z) - 1, \nn\\
    \phi(z) &=&
    \left\{
        \begin{array}{ll}
            \frac{1}{4} \brs{\pi^2 - \ln^2 \frac{1+\sqrt{1-z}}{1-\sqrt{1-z}}} +
            i \frac{\pi}{2} \ln \frac{1+\sqrt{1-z}}{1-\sqrt{1-z}}, & z < 1, \\
            \br{\arctan \frac{1}{\sqrt{z-1}}}^2, & z>1. \\
        \end{array}
    \right.
\ea
We remind that for the quark loop contribution
the imaginary part of the function $\Phi(z)$ must be omitted
(see \cite{Pervushin:1985yi}),
and for the meson loop contribution both the real and
possible imaginary parts are relevant.

Similar expressions were obtained in \cite{Ebert:1996pc},
where the imaginary part of the quark loop contribution was
taken into account.


The vertices of the quark-meson and quark-photon interactions
were given in (\ref{QuarkMesonLagrangian}).
The vertices of the meson-meson interaction in the framework of
the NJL model have the form (for details see \cite{Volkov:1986zb,Ebert:1994mf,Bystritskiy:2007wq}):
\ba
    V_{\sigma_s \pi^+\pi^-} &=& V_{a_0 \pi^+\pi^-} = 0, \nn\\
    V_{\sigma_u K^+K^-} &=& V_{a_0 K^+K^-} = -2\br{2m_u - m_s} \frac{g_K^2}{g_{\sigma_u}},
    \label{MesonMesonVertexes}\\
    V_{\sigma_s K^+K^-} &=& 2\sqrt{2}\br{2m_s - m_u} \frac{g_K^2}{g_{\sigma_s}}, \nn\\
    V_{\sigma_u \pi^+\pi^-} &=& -4m_u \frac{g_\pi^2}{g_{\sigma_u}}. \nn
\ea
The general structure of the two-photon scalar meson decay amplitudes has the form
\ba
T_{S\gamma\gamma}=-\frac{\alpha g_{\sigma_u}}{\pi m_u}\br{g_{\mu\nu}\br{q_1 q_2} - q_1^\nu q_2^\mu}a_{S\gamma\gamma}.
\ea
The expression for the width has the form:
\ba
\Gamma_{S\gamma\gamma}=\frac{M_S^3}{64\pi}\frac{\alpha^2g^2_{\sigma_u}}{\pi^2m_u^2}|a_{S\gamma\gamma}|^2.
\ea

\subsection{The decay $a_0\to 2\gamma$}

The amplitude $a_{a_0\gamma\gamma}$ of $a_0\to \gamma\gamma$ contains the contribution of
$u$, $d$ quarks and the $K$-meson intermediate states.
The color-charge factor associated with $u$, $d$ quarks is $N_c\br{\frac{4}{9}-\frac{1}{9}}=1$.
Thus,
\ba
a^{u,d}_{a_0\gamma\gamma} &=&
 F\br{z_{a_0}^u}.
\ea
Taking the $K$-meson loop contribution we obtain:
\ba
a_{a_0\gamma\gamma} &=&
    a^{u,d}_{a_0\gamma\gamma} + a^{K}_{a_0\gamma\gamma} = \nn\\
    &=&
    F\br{z_{a_0}^u}
    -
    \frac{m_u}{g_{\sigma_u}}
    \frac{2\br{2m_u-m_s} g_K^2}{4 g_{\sigma_u} M_K^2}
    z_{a_0}^K \Phi\br{z_{a_0}^K}=\nn\\
    &=&0.482-0.114=0.367.
\ea
The corresponding width is
\ba
\Gamma_{a_0(980)\to\gamma\gamma} &=& \br{2.25\KeV} \brm{a_{a_0\gamma\gamma}}^2 = 0.29\KeV. \nn
\ea
This value is in satisfactory agreement with the experimental values
(see Table~\ref{TableOfDecays}).

\subsection{The decay $f_0(980)\to \gamma\gamma$}

In the case of the $f_0\to \gamma\gamma$ decay we also have the contribution of
$u$, $d$ and $s$ quarks and the $K$-meson and pion intermediate states.
We should recall that $f_0(980)$ meson consists of two components:
$\sigma_u$ and $\sigma_s$ (see (\ref{Mixing})).
The color-charge factor associated with $u$, $d$ quarks is $N_c\br{\frac{4}{9}+\frac{1}{9}}=\frac{5}{3}$
for $\sigma_u$-component of $f_0$ and $N_c\frac{1}{9}=\frac{1}{3}$
for $\sigma_s$-component of $f_0$.
Taking the $K$-meson and the $\pi$-meson loop contribution we obtain
\ba
a_{f_0\to \gamma\gamma} &=&
\frac{5}{3}F\br{z_{f_0}^u} \sin\alpha
-
\frac{\sqrt{2}}{3}F\br{z_{f_0}^s}\br{\frac{g_{\sigma_s} m_u}{g_{\sigma_u} m_s}} \cos\alpha
+
\nn\\
&+&
\left(
    -\frac{g_K^2}{g_{\sigma_u}^2}
    \frac{m_u}{4 M_K^2}
    2\br{2 m_u-m_s}\sin\alpha
\right.
+\nn\\
&&\qquad+
\left.
    \frac{g_K^2}{g_{\sigma_u}g_{\sigma_s}}
    \frac{m_u}{4 M_K^2}
    2\sqrt{2}\br{2 m_s-m_u}\cos\alpha
\right)
z_{f_0}^K \Phi\br{z_{f_0}^K}
- \nn\\
&-&
    \sin\alpha
    \frac{m_u^2}{M_\pi^2}
    \frac{g_\pi^2}{g_{\sigma_u}^2}
    z_{f_0}^\pi \Phi\br{z_{f_0}^\pi}=\nn\\
&=& 0.157-0.417-0.022+0.589+0.082-0.038 i= \nn\\
&=& 0.385-0.038 i.
\ea
For the width we have
\ba
\Gamma_{f_0(980)\to\gamma\gamma} &=& \br{2.25\KeV} \brm{a_{f_0\gamma\gamma}}^2 = 0.33\KeV. \nn
\ea
This value is also in satisfactory agreement with the experimental values
(see Table~\ref{TableOfDecays}).

\subsection{The decay $\sigma\to \gamma\gamma$}

In the case of the $\sigma\to \gamma\gamma$ decay we have
\ba
a_{\sigma\to \gamma\gamma} &=&
\frac{5}{3}F\br{z_{\sigma}^u} \cos\alpha
+
\frac{\sqrt{2}}{3}F\br{z_{\sigma}^s}\br{\frac{g_{\sigma_s} m_u}{g_{\sigma_u} m_s}} \sin\alpha
\nn\\
&-&
\left(
    \frac{g_K^2}{g_{\sigma_u}^2}
    \frac{m_u}{4 M_K^2}
    2\br{2 m_u-m_s}\cos\alpha
\right.
+\nn\\
&&\qquad+
\left.
    \frac{g_K^2}{g_{\sigma_u}g_{\sigma_s}}
    \frac{m_u}{4 M_K^2}
    2\sqrt{2}\br{2 m_s-m_u}\sin\alpha
\right)
z_{\sigma}^K \Phi\br{z_{\sigma}^K}
-\nn\\
&-&
    \cos\alpha
    \frac{m_u^2}{M_\pi^2}
    \frac{g_\pi^2}{g_{\sigma_u}^2}
    z_{\sigma}^\pi \Phi\br{z_{\sigma}^\pi}=
\nn\\
&=& 1.89+0.057-0.041-0.043+0.92-0.98 i=
\nn\\
&=& 2.78 - 0.98 i.
\ea
Let us notice that unlike $f_0(980)\to 2\gamma$ decay,
where contributions of $u$ and $d$ quarks have opposite signs,
here they have the same sign.

The corresponding width is
\ba
\Gamma_{\sigma(600)\to\gamma\gamma} &=& \br{0.51\KeV} \brm{a_{\sigma\gamma\gamma}}^2 = 4.3\KeV.
\ea

The experimental value of the mass and the width of the $\sigma$ meson is not well established.
We present the width of $\sigma$ for two other masses: $M_\sigma=450\MeV$ and $M_\sigma=550\MeV$.
They are
\ba
\Gamma_{\sigma(450)\to\gamma\gamma}= 2.18\KeV,\nn \\
\Gamma_{\sigma(550)\to\gamma\gamma}= 3.53\KeV.\nn
\ea
The comparison of our results with the experimental data and some other model
predictions is given in Table~\ref{TableOfDecays}.

\section{The decays $a_0\to\omega(\rho)\gamma$, $f_0\to\omega(\rho)\gamma$}
\label{ChapterNew}

In this section, we will consider the following decays
\footnote{
Let us notice that in \cite{Radzhabov:2007wk}
the decays $\rho(\omega) \to 2\pi \gamma$ were considered.
The $\sigma$ meson intermediate state was taken into account.
}
:
\ba
a_0(p) &\to& \omega(q) + \gamma(k_1), \nn \\
a_0(p) &\to& \rho(q) + \gamma(k_1), \nn \\
f_0(p) &\to& \omega(q) + \gamma(k_1), \nn \\
f_0(p) &\to& \rho(q) + \gamma(k_1), \nn
\ea
\ba
p^2=M_S^2, \qquad q^2=M_V^2, \qquad k_1^2=0,
\ea
where $M_S=M_{f_0,a_0}=980\MeV$ is the mass of decaying scalar meson
\cite{Amsler:2008zzb} and
$M_V=M_{\omega,\rho}$ is the mass of vector meson.
Now the matrix element in the general case has the form:
\ba
    M_i&=&e~\frac{g_\rho}{2} A_i \br{q_\nu k_{1\mu}-g_{\mu\nu}\br{qk_1}} e_\gamma^\nu e_V^\mu, \\
    &&e_\gamma = e\br{k_1}, \qquad e_V = e\br{q},
\label{AmpView}
\ea
where $i=\brf{a_0\to\omega\gamma, a_0\to\rho\gamma,f_0\to\omega\gamma,f_0\to\rho\gamma}$,
and $A_i$ contains all vertex constants and dynamic information of the processes.
The radiative decay width then has the form:
\ba
\Gamma_i=\frac{\alpha\br{M_S^2-M_V^2}^3}{32 M_S^3}g_\rho^2\brm{A_i}^2.
\ea
To calculate quark and meson loop contributions to coefficients $A_i$,
we need quark-meson (see (\ref{QuarkMesonLagrangian})) and
meson-meson vertices (see (\ref{MesonMesonVertexes})),
and vertices of interaction of vector mesons with the pseudoscalar ones
\cite{Volkov:1986zb,Volkov:1993jw,Ebert:1994mf}:
\ba
g_{\omega^\mu K^+ K^-} &=& g_{\rho^\mu K^+ K^-} = \frac{g_\rho}{2} \br{p_+ - p_-}^\mu, \nn\\
g_{\rho^\mu \pi^+ \pi^-} &=& g_\rho \br{p_+ - p_-}^\mu. \nn
\ea

\subsection{The decays $a_0\to\omega\gamma$, $a_0\to\rho\gamma$}

Let us consider the decay of the isoscalar meson $a_0$ with $\omega$-meson production.
The amplitude of this decay will contain contributions from the quark loop and
$K$-meson loop. The $\pi$-meson loop is forbidden since the $a_0\to 2\pi$ vertex is absent.

Quark contribution consists of $u$ and $d$ quark loops (see Fig.~\ref{Fig2},~a)
(since $\omega$ does not contain $s$ quarks):
\begin{figure}
\includegraphics[width=0.8\textwidth]{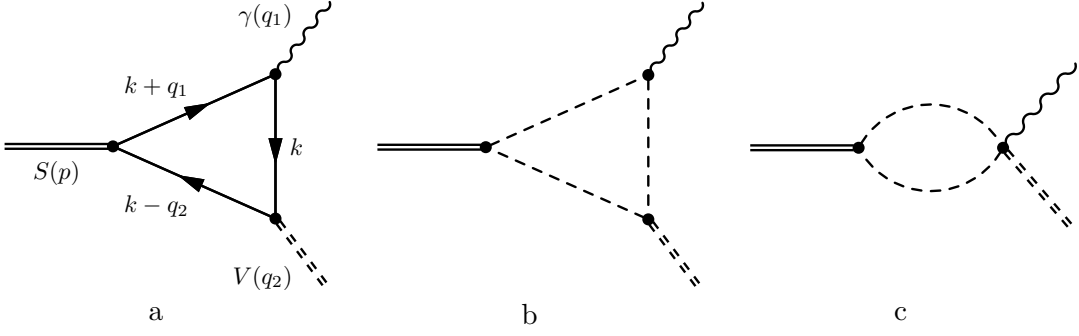}
\caption{The Feynman diagrams of quark and meson contributions
to decays of scalar mesons to photon and vector meson: $S \to \gamma V$.
\label{Fig2}}
\end{figure}
\ba
M^{(u,d)}_{a_0\to\omega\gamma} &=& e~\frac{g_\rho}{2}C^{(u,d)}_{a_0\to\omega\gamma}
\int \frac{d^4 k}{i\pi^2}
\frac{Sp\brs{\br{\hat q+\hat k+m_u}\br{\hat k-\hat k_1+m_u} \hat e_\gamma \br{\hat k+m_u} \hat e_\omega}}
{\br{\br{q+k}^2 - m_u^2}\br{k^2 - m_u^2}\br{\br{k-k_1}^2 - m_u^2}},
\ea
where $C^{(u,d)}_{a_0\to\omega\gamma}=3 g_{\sigma_u}$, where
factor $3=N_c$ is the color factor.

Standard Feynman procedure of denominators joining and loop momenta
integration leads to:
\ba
M^{(u,d)}_{a_0\to\omega\gamma} &=& e~\frac{g_\rho}{2}C^{(u,d)}_{a_0\to\omega\gamma} \mbox{Re}\br{I_u}
\br{q_\nu k_{1\mu}-g_{\mu\nu}\br{qk_1}} e_\gamma^\nu e_\omega^\mu,
\ea
where
\ba
I_u &=&
4 m_u \int\limits_0^1 dx \int\limits_0^1 dy
\frac{y\br{1-4y^2 x\br{1-x}}}{m_u^2 - y(1-y)(1-x)q^2-x(1-x)y^2p^2+i\epsilon}.
\ea
Let us consider now kaon loop contributions.
The kaon contribution consists of diagrams of two types (see Fig.~\ref{Fig2}~b, c).
This contribution can be written in a form similar to (\ref{AmpView}):
\ba
M^{(K)}_{a_0\to\omega\gamma} &=& e~\frac{g_\rho}{2}C^{(K)}_{a_0\to\omega\gamma} I_K
\br{q_\nu k_{1\mu}-g_{\mu\nu}\br{qk_1}} e_\gamma^\nu e_\omega^\mu,
\ea
where $C^{(K)}_{a_0\to\omega\gamma}=g_{a_0K^+K^-}$ and
\ba
I_K &=&
\int\limits_0^1 dx \int\limits_0^1 dy
\frac{4y^2x(1-x)y}{M_K^2 - y(1-y)(1-x)q^2-x(1-x)y^2p^2+i\epsilon}.
\label{IK}
\ea
More details of this type of diagram calculation can be found in \cite{Bystritskiy:2007wq}.

The amplitude of the process $a_0\to\omega\gamma$ then has the form
\ba
M_{a_0\to\omega\gamma} = e~A_{a_0\to\omega\gamma}
\br{q_\nu k_{1\mu}-g_{\mu\nu}\br{qk_1}} e_\gamma^\nu e_\omega^\mu,
\ea
where
\ba
A_{a_0\to\omega\gamma} = 3 g_{\sigma_u} \mbox{Re}\br{I_u} + g_{a_0 K^+ K^-} I_K
=
-1.78374 + 0.159415 = -1.62433.
\ea
The decay width is:
\ba
    \Gamma_{a_0\to\omega\gamma} &=& 115\KeV. \nn
\ea

The decay $a_0\to\rho\gamma$ can be considered in complete analogy with the
decay $a_0\to\omega\gamma$. Let us note that the quark contribution
to $a_0\to\rho\gamma$ is three times smaller than the
quark contribution to the $a_0\to\omega\gamma$ decay.
However, the kaon loop contributions are the same.
As a result, the amplitude of the $a_0\to\rho\gamma$ decay
has the form:
\ba
M_{a_0\to\rho\gamma} &=& e~\frac{g_\rho}{2}A_{a_0\to\rho\gamma}
\br{q_\nu k_{1\mu}-g_{\mu\nu}\br{qk_1}} e_\gamma^\nu e_\rho^\mu, \\
A_{a_0\to\rho\gamma} &=& g_{\sigma_u} \mbox{Re}\br{I_u} + g_{a_0 K^+ K^-} I_K
=
-0.598209 + 0.156921. \nn
\ea
The decay width is
\ba
    \Gamma_{a_0\to\rho\gamma} &=& 8.5\KeV. \nn
\ea
Let us note that in both these processes the main contribution comes from
quark loops.

\subsection{The decays $f_0\to \omega\gamma$ and $f_0\to \rho\gamma$}

The total amplitude for the $f_0\to \omega\gamma$ decay
has the form:
\ba
A_{f_0 \to \omega\gamma} &=&
\sin\alpha~A_{\sigma_u \to \omega\gamma}^{u,d} +
\sin\alpha~A_{\sigma_u \to \omega\gamma}^{K} +
\cos\alpha~A_{\sigma_s \to \omega\gamma}^{K}
=
-0.115 + 0.0195 + (-1.068)
=
-1.1635.
\ea

The decay width is then:
\ba
    \Gamma_{f_0\to\omega\gamma} &=& 60\KeV. \nn
\ea

Let us consider the decay $f_0\to \rho\gamma$.
The quark loop contribution to $\sigma_u$ component
here exceeds the relevant component of the $f_0\to \omega\gamma$ decay:
\ba
A_{f_0 \to \rho\gamma} &=&
\sin\alpha~A_{\sigma_u \to \rho\gamma}^{\br{u,d}}
+
\sin\alpha~A_{\sigma_u \to \rho\gamma}^{\pi}
+
\sin\alpha~A_{\sigma_u \to \rho\gamma}^{K}
+
\cos\alpha~A_{\sigma_s \to \rho\gamma}^{K}
= \nn\\
&=&
-0.353 + \br{0.0436 - 0.00542 i} + 0.027 + \br{-1.05}
=
-1.3324 - 0.00542 i.
\ea
The $f_0\to \rho\gamma$ decay width is then:
\ba
    \Gamma_{f_0\to\rho\gamma} &=& 82\KeV. \nn
\ea
Note that in both the cases (the decays of $f_0(980)$ to $\rho\gamma$ or $\omega\gamma$)
the main contribution comes from the kaon loops related with the $\sigma_s$ component.

Unfortunately, at present we do not have any experimental data for this
decays.

\section{Conclusion}
\label{Conclusion}

The calculations of the radiative decays in the NJL model show an important
role of both the quark
and meson loops. Moreover, for the $f_0$ meson decay the kaon loop turns out to
provide the dominant contribution.
This can be understood if one takes into account that in the $2\gamma$
decays the fractional quark charge gives effectively
factor $1/N_c^2$ into quark loop contribution.
It is worth noticing that the situation here is
similar to the one that takes place in the case of
$\phi\to f_0 \gamma$ decay \cite{Bystritskiy:2007wq}
and $a_0\to\rho(\omega)\gamma$ and $f_0\to\rho(\omega)\gamma$.
This fact permits one to understand the success
of such models as the model of a kaon molecule \cite{Weinstein:1990gu,Branz:2008ha}
as well as the four-quark model \cite{Achasov:1987ts,Achasov:2008ut}.

The description of radiative scalar isoscalar meson $f_0$ decays
was also considered
in the linear $\sigma$-model \cite{Kleefeld:2001ds,vanBeveren:2002mc,vanBeveren:2008st}.
However, in these papers the scalar meson $a_0(980)$ and $\sigma$
radiative decays were not considered.

The NJL model used here allows us to take into account
both the quark-antiquark state which manifests itself in the form of quark loops,
and the hidden four-quark state which shows up as meson loops.
Let us emphasize that in the framework of the standard NJL model we can describe the
radiative decays without any additional parameters.

The results obtain in this paper (see Table~\ref{TableOfDecays})
are in agreement with the existing experimental data.
Unfortunately, the data for the $a_0\to \rho(\omega)\gamma$ and $f_0\to \rho(\omega)\gamma$
decays are absent up to now and our results for these decays can be considered
as a prediction.


\begin{acknowledgments}
The authors wish to thank Prof. N. N. Achasov, Prof. S. B. Gerasimov
and Prof. V. N. Pervushin for fruitful discussions.
We also acknowledge the support of INTAS grant no. 05-1000008-8528.
\end{acknowledgments}

\begin{table}[p]
\caption{The table of two-gamma decays of the scalar mesons
$\sigma(600)$, $f_0(980)$ and $a_0(980)$.
\label{TableOfDecays}}
\begin{tabular}{|l|l|}
\hline
\hline
$\Gamma\br{a_0\to\gamma\gamma}\br{exp.}$, $\KeV$ &
$\Gamma\br{a_0\to\gamma\gamma}\br{theor.}$, $\KeV$ \\
\hline
\hline
$0.30\pm 0.10$ \cite{Amsler:1997up} & $0.29$ (This paper)\\
\hline
\hline
$\Gamma\br{f_0\to\gamma\gamma}\br{exp.}$, $\KeV$ &
$\Gamma\br{f_0\to\gamma\gamma}\br{theor.}$, $\KeV$ \\
\hline
\hline
$0.42$ (Solution A) \cite{Pennington:2008xd} & $0.33$ (This paper) \\
\hline
$0.10$ (Solution B) \cite{Pennington:2008xd} & $0.21-0.26$ \cite{Branz:2008ha} \\
\hline
$0.205^{+0.095+0.147}_{-0.083-0.117}$ \cite{Mori:2006jj} & $0.22$ \cite{Hanhart:2007wa} \\
\hline
$0.28^{+0.09}_{-0.13}$ \cite{Boglione:1998rw} & $0.33$ \cite{Schumacher:2006cy} \\
\hline
$0.42 \pm 0.06 \pm 0.18$ \cite{Oest:1990ki} & $0.31$ \cite{Scadron:2003yg} \\
\hline
$0.29 \pm 0.07 \pm 0.12$ \cite{Boyer:1990vu} & $0.28^{+0.09}_{-0.13}$ \cite{Anisovich:2001zp} \\
\hline
$0.31 \pm 0.14 \pm 0.09$ \cite{Marsiske:1990hx} & $0.20$ \cite{Oller:1997yg} \\
\hline
$0.63 \pm 0.14$ \cite{Morgan:1990kw} & $0.24$ \cite{Efimov:1993zg} \\
\hline
                                     & $0.27$ \cite{Achasov:1981kh} \\
\hline
\hline
$\Gamma\br{\sigma\to\gamma\gamma}\br{exp.}$, $\KeV$ &
$\Gamma\br{\sigma\to\gamma\gamma}\br{theor.}$, $\KeV$ \\
\hline
\hline
$3.1 \pm 0.5$ (Solution A) \cite{Pennington:2008xd} & $4.30$ ($M_\sigma=600\MeV$)(This paper)\\
\hline
$2.4 \pm 0.4$ (Solution B) \cite{Pennington:2008xd} & $3.53$ ($M_\sigma=550\MeV$)(This paper)\\
\hline
$4.1 \pm 0.3$ \cite{Pennington:2006dg} & $2.18$ ($M_\sigma=450\MeV$)(This paper)\\
\hline
$3.8 \pm 1.5$ \cite{Boglione:1998rw} & \\
\hline
$5.4 \pm 2.3$ \cite{Morgan:1990kw} & \\
\hline
$10 \pm 6$ \cite{Courau:1986gn} & \\
\hline
\end{tabular}
\end{table}

%
%

\end{document}